# A Study of Energy Transfer of Wind and Ocean Waves

*Mason Bray*[1] *and Sharhdad Sajjadi*[2]

[1]*MA* 490 *Section* 01 *Embry – Riddle Aeronautical University Daytona Beach, FL*
[2]*Department of Mathematics, Embry – Riddle Aeronautical University Daytona Beach, FL*

(Dated: December 2013)

**Abstract.** To develop a better understanding of energy transfer between wind and different types of waves a model was created to determine growth factors and energy transfers on breaking waves and the resulting velocity vectors. This model was used to build on the research of Sajjadi *et al*(1996) on the growth of waves by sheared flow and takes models of wave velocities developed by Weber and Melsom(1993) and end energy transfer by Sajjadi, Hunt and Drullion(2012).

## I.    Introduction

The topic of atmospheric oceanic interactions in tropical systems is very important issue to the meteorological community as well as society as whole. With the increase in number of tropical cyclones per year increasing over the last decade the understanding of the development of all variables that may intensify tropical systems has become paramount.

The interactions between the atmosphere and ocean waves has been one of the cornerstones of tropical cyclone research since the introduction of the Wind Induced Surface Heat Exchange, or WISHE, model introduced by Yano and Emanuel (1991) as well as presented by Emanuel (1994) which introduced the theory of a positive feedback loop between wind speed evaporation and intensification of the system. The logic behind this model is that as wind blows over the waves, the waves grow and become steeper. At some point the waves reach a critical steepness and break. This event releases spray that can evaporate and increase the energy of the system. The growth of the waves can be expressed through an energy transfer parameter introduced in Sajjadi, Hunt and Drullion (2012) and is used to determine wave speeds from a model developed by Weber and Melsom (1993).

The study of the topic of this paper is divided into three stages. First an understanding of the energy transfer from wind to wave must be established. Second the growth of the waves must be parameterized and finally the velocity vector can be determined.

Before an energy parameter can be established how waves can move and grow. Wave speed can contain two components, $c_r$ the real component of the wave speed which controls propagation, and $c_i$ which is the wave growth consisting of the imaginary part of wave speed [Sajjadi, Hunt, and

Drullion (2012)]. To produce a coordinate that can be used in conjunction with energy flux $c_r$ we can take the ratio of the real part of wave speed and the frictional velocity $U_*$ to create the wave age $\frac{c_r}{U_*}$.

To be able to model the total energy flux the nature of energy transfer must be understood. The total energy transfer, $\beta$, is made of two parts, $\beta_c$ the energy transfer parameter of the critical layer, and $\beta_t$ the energy transfer parameter due to turbulence. $\beta$ can be expressed as a function of the wave age which can be modeled by the energy transfer equation from Sajjadi, Hunt and Drullion (2012).

The second step of this study is to determine the growth rate of waves. As previously stated waves grow if the imaginary component of wave speed is nonzero and can be written as a function of wavelength. Total wave growth $\zeta$, can be represented as a wave growth of the air above the wave, $\zeta_a$ as well as the wave growth of the water, $\zeta_w$. Finally velocities can be modeled at fixed energy transfer parameters as a function of wave displacement, x, and time.

## II. Development of Equations

In order to develop our model three existing models were used and their results combined to create velocity vectors as a result. The first equation was developed by Sajjadi, Hunt, and Drullion as a parameterization of $\beta_c$ as a function of wave age which is given by

$$\beta_c = \pi \xi_c L_0^4 [1 + (4 - \frac{1}{3}\pi^2 + \hat{c}_i^2)\Lambda^2 \tag{1.1}$$

Noting that

$$L_0 = \gamma - \ln(2\xi_c) = \Lambda^{-1}$$

$$\hat{c}_i = \kappa \frac{c_r}{10 U_*}$$

$$\xi_c = k z_c = \Omega \left(\frac{U_*}{c_r \kappa}\right)^2 e^{\frac{kc_r}{U_*}}$$

Where $\gamma = 0.5772$ is Euler's constant, $\kappa = .04$ is von Karman's constant, and $\Omega$ is given as $1.25 x 10^{-2}$.

The energy transfer parameter due to turbulence, $\beta_t$, is developed by Sajjadi, Hunt, and Drullion and given by the equation.

$$\beta_t = 5\kappa^2 L_0 \tag{1.2}$$

Combining equation (1.1) and (1.2) results in the total energy transfer parameter. For this study we used $\beta$ as a function of $\left(\dfrac{c_r}{U_*}\right)$.

To develop the equation for $\zeta$ we must first develop the equation for the imaginary part of wave speed in air and water; In this study we use equations for $c_i$ developed by Sajjadi Hunt and Drullion (2012)

So that $c_i$ for air can be expressed as

$$c_{i\_a} = sU_* \left\{ \left(\frac{0.322}{w^2}\right)\left(\frac{U_*}{k\upsilon_a}\right)^{\frac{2}{3}}\left(\frac{c_r}{U_*}\right) + \left(\frac{.343}{w}\right)\left(\frac{U_*}{k\upsilon_a}\right)^{\frac{1}{3}} \right\} \tag{1.3}$$

Where s is the ratio of the densities of air and water which is equal to $\dfrac{1}{1000}$, w = 2.3, $\upsilon_a$ is the kinematic viscosity of air $1.5 \times 10^{-5}\, m^2/s$, and k is the wave number which is a function of wavelength, and thus we can represent the wave growth rate of air to be

$$\zeta_a = kc_{i\_a} \tag{1.4}$$

And the wave growth rate of water can be expressed as

$$\zeta_w = 2k^2 \upsilon_w \tag{1.5}$$

Where $\upsilon_w$ is the kinematic viscosity of water, $0.8 \times 10^{-6}\, m^2/s$. Thus the total wave growth rate can be represented as the sum of equations (1.4) and (1.5).

To develop the model for velocities the equations for each velocity component u, v, and w were formed from equations in Weber and Hunt (2993) which can be expressed as

$$\begin{aligned}
u &= \omega\zeta_0 e^{\beta t}\left[\cos(kx-\omega t) - \frac{\beta}{\omega}\sin(kx-\omega t)\right]e^{kz} \\
v &= f\zeta_0 e^{\beta t}\sin(kx-\omega t)e^{kz} \\
w &= \omega\zeta_0 e^{\beta t}\left[\sin(kx-\omega t) + \frac{\beta}{\omega}\cos(kx-\omega t)\right]e^{kz}
\end{aligned} \tag{1.6}$$

### III. Results and Discussion

To determine the correlation between energy transfers to waves $\beta, \beta_c, and \beta_t$ were plotted against wave age

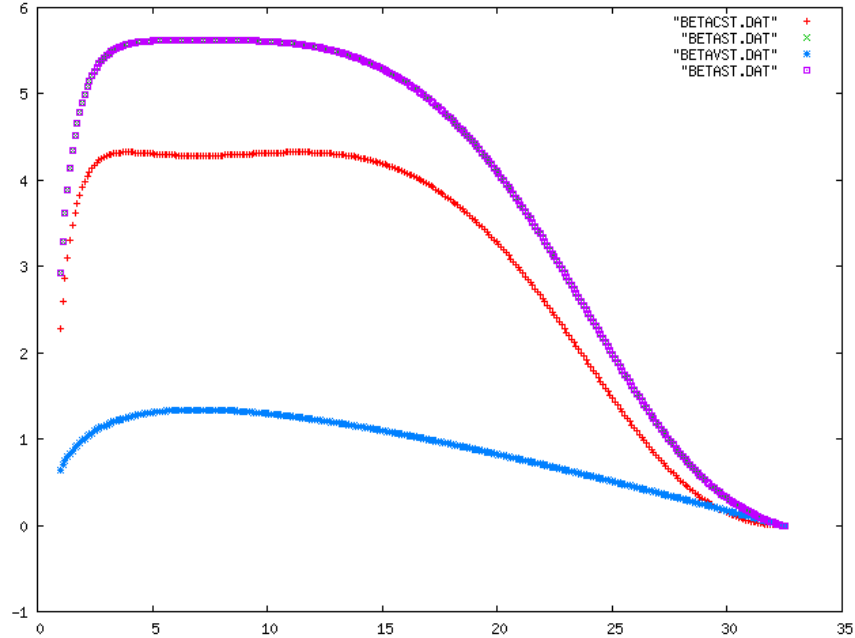

Which compares well to Sajjadi(1996) The x axis represents $\left(c_r/U_*\right)$ while $\beta, \beta_c, and \beta_t$ are represented on the y axis. This table shows that energy transfer occurs early and then tapers off as wave age increases towards 32.

Wave growth was also plotted versus wavelength for various frictional velocities

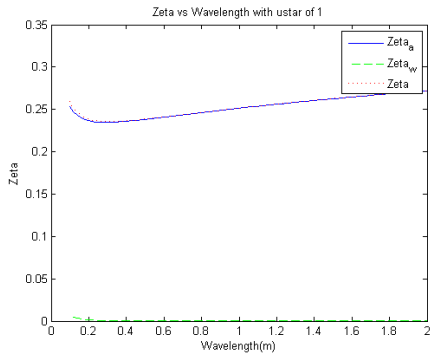 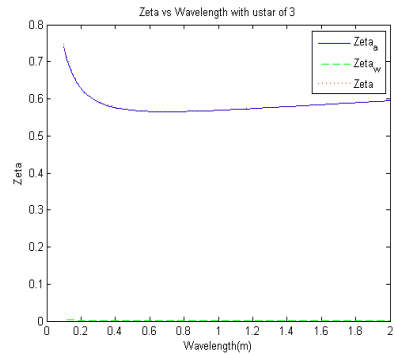

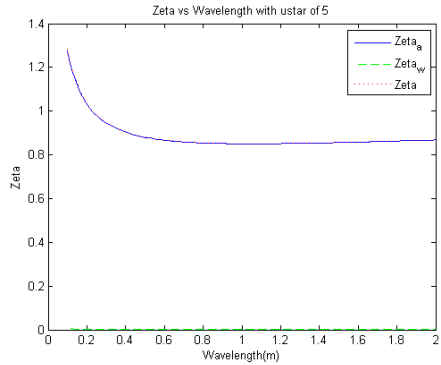
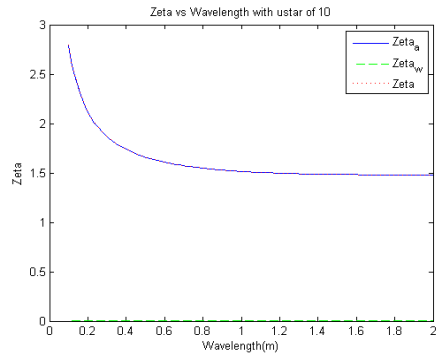
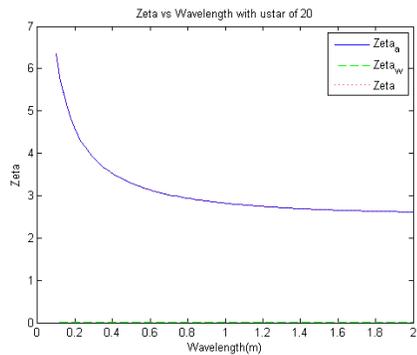
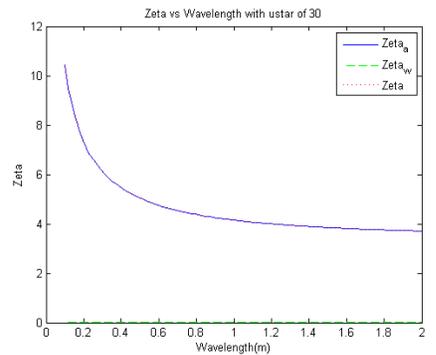

The common trend between all of the figures above is how much more of a factor $\zeta_a$ is in the total wave growth than $\zeta_w$. Wave growth also a function of the imaginary part of wave speed and any change in $c_i$ will have an effect on $\zeta$. The figure below represents $c_i$ vs $\left(c_r/U_*\right)$

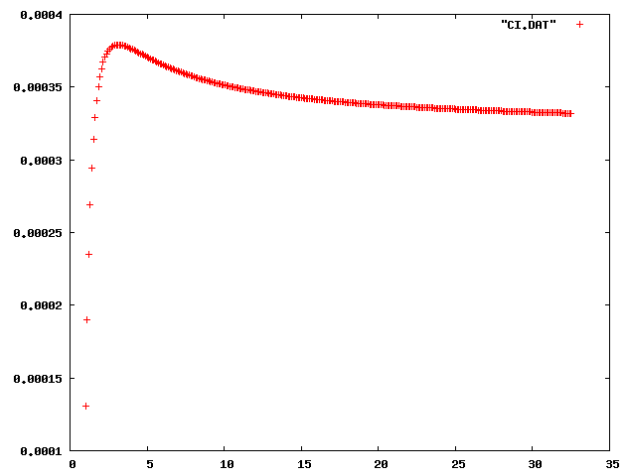

The imaginary part of wave speed is represented by the y axis and wave age on the x axis. It can be seen that a maximum $c_i$ occurs at a relatively young wave age and then it slowly tapers off as it approaches 32.

The final part of the study involved plotting velocities in respect to x and t. This was done for two different wave growth values one calculated by Weber and Melsom (1993) and the other by Sajjadi Hunt and Drullion.

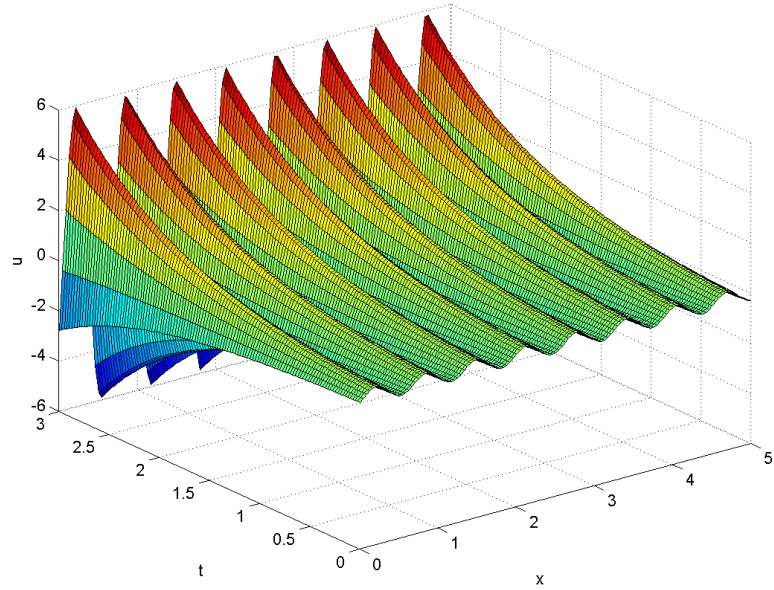

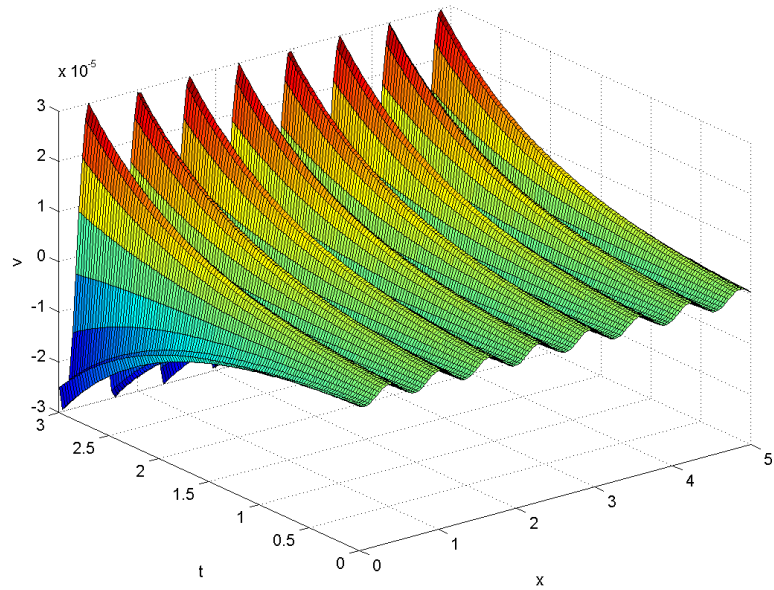

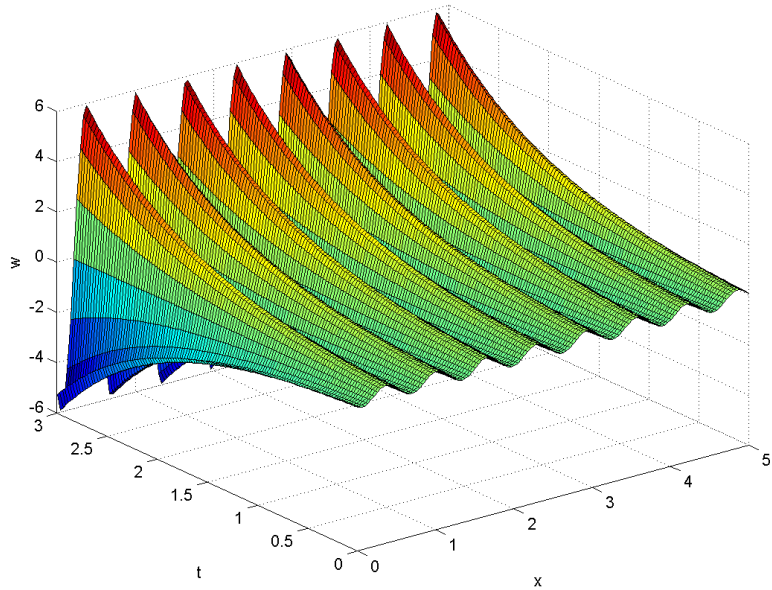

The above three pictures were calculated using Weber and Melsom's (1993) growth rate $\beta = kc_i$. It should be noted that velocity is sinusoidal in x and increases as time increases. A cross section of each was plotted as well with x held constant at three different points.

At x = 0

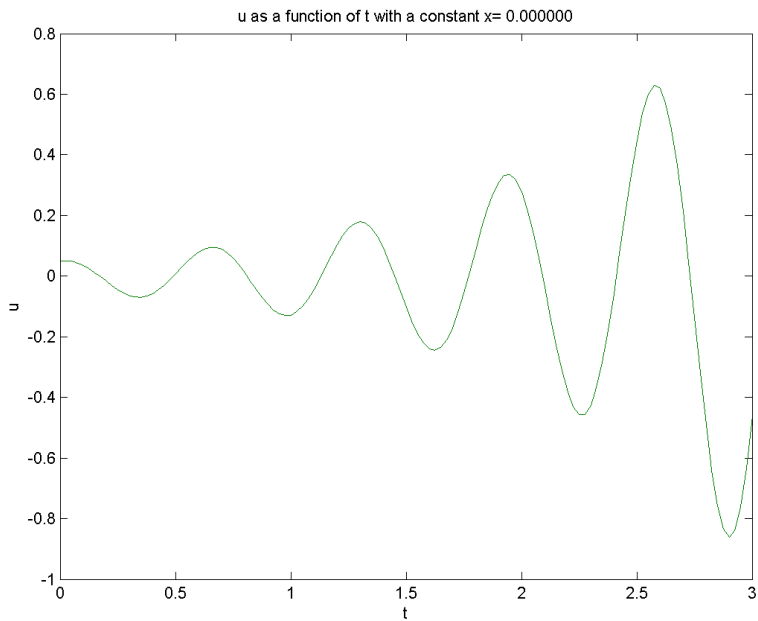

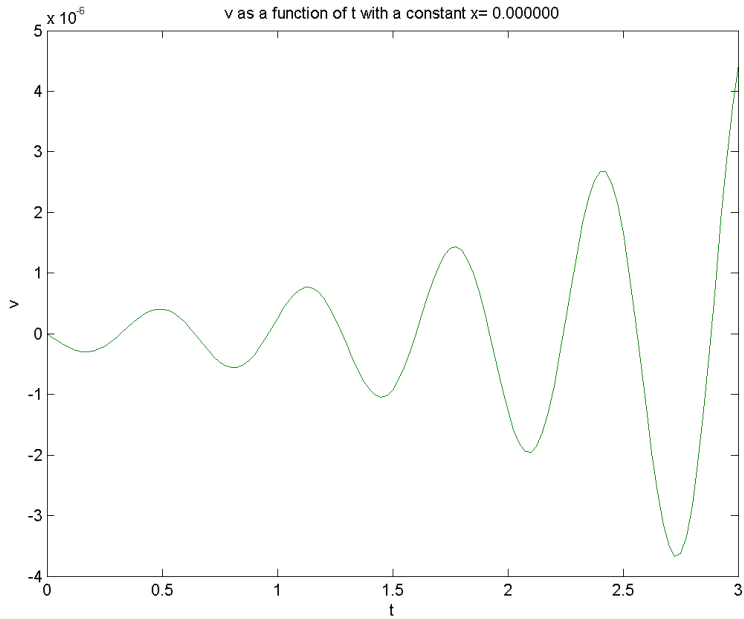

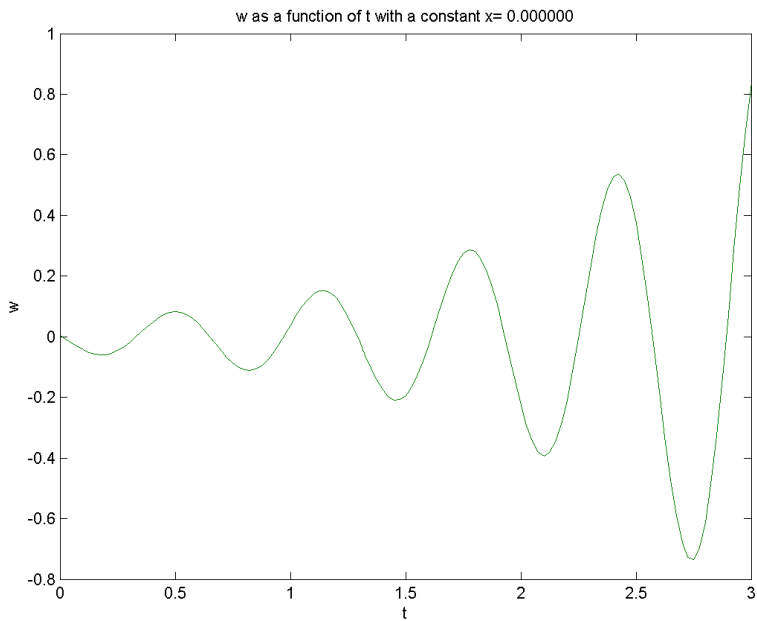

All velocities are sinusoidal at the given x, but not that the magnitude of the meridianal velocity, v, is much smaller than the magnitudes of the zonal and vertical velocities. This is due to it being affected by Coriolis force. The figures for $x = \frac{\pi}{2}$ and $x = \pi$ are similar with magnitudes only changing.

$X = \dfrac{\pi}{2}$

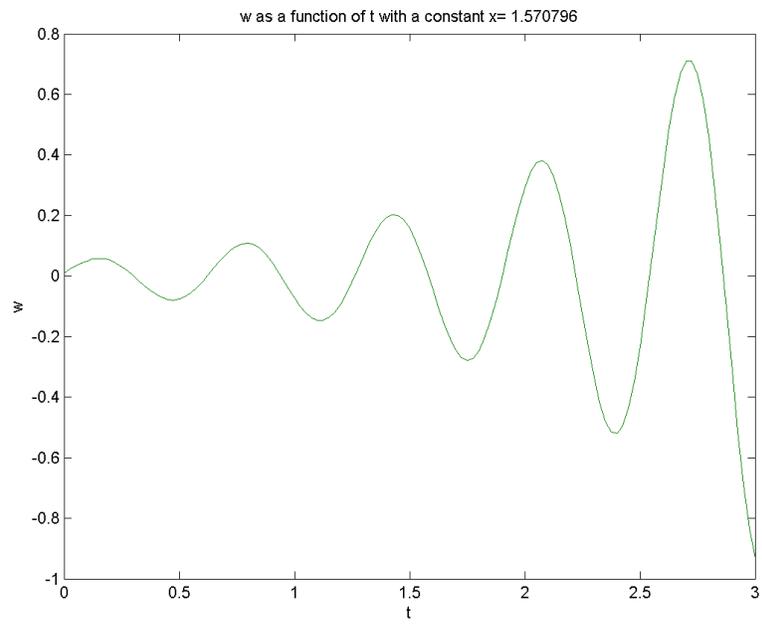

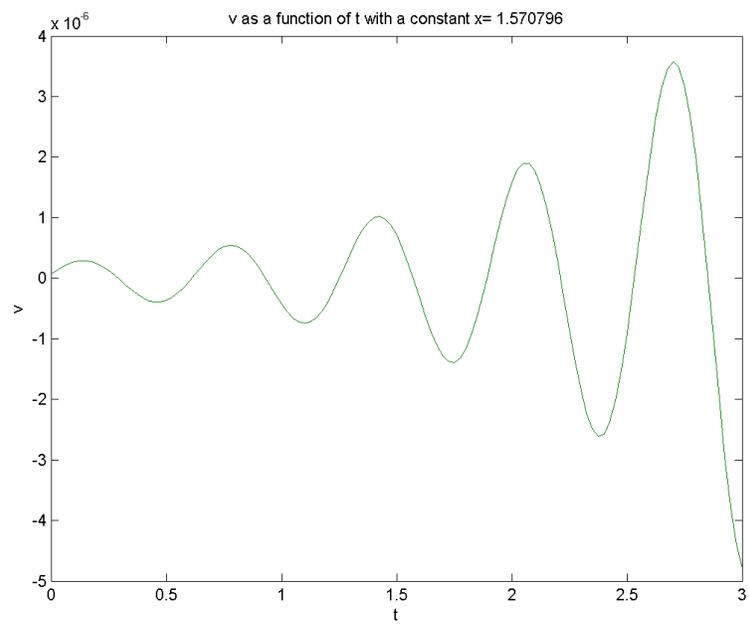

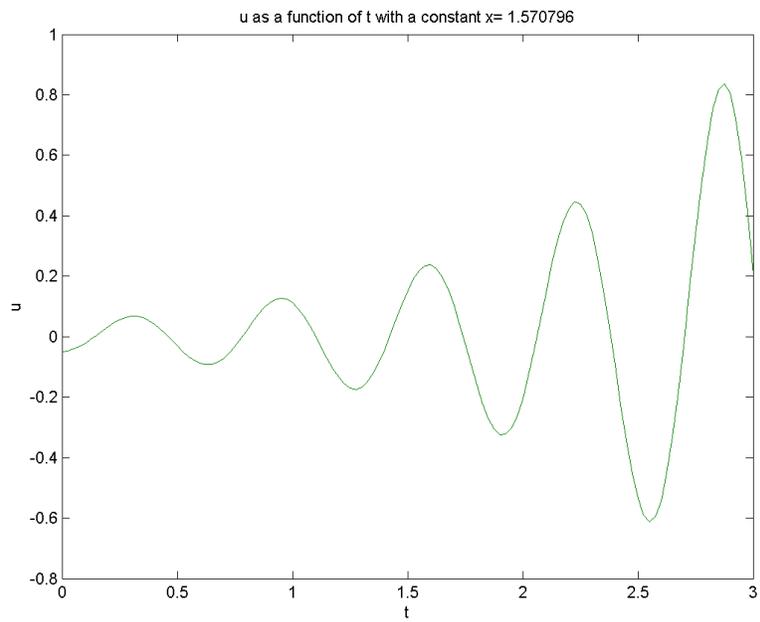

X = π

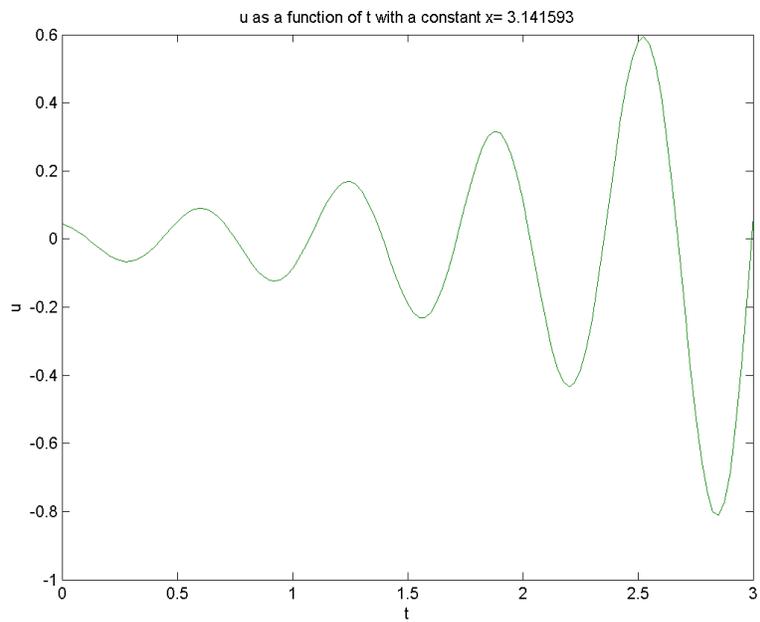

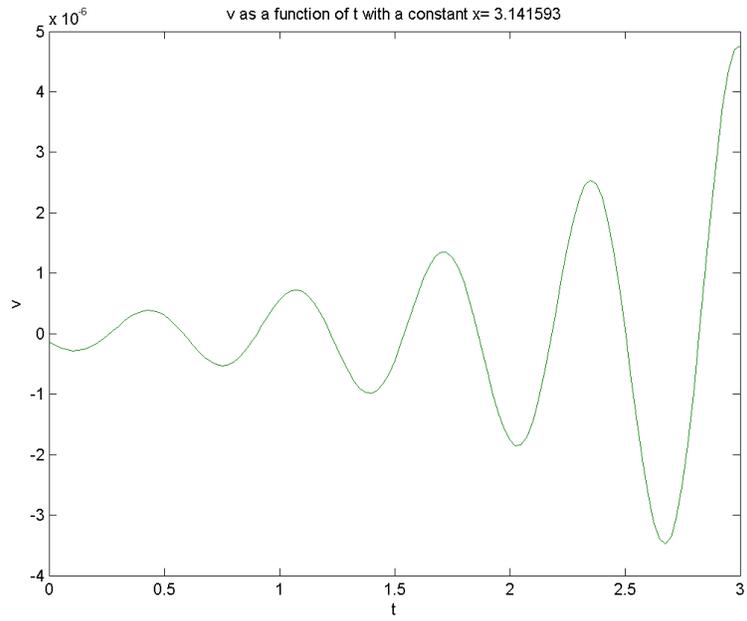
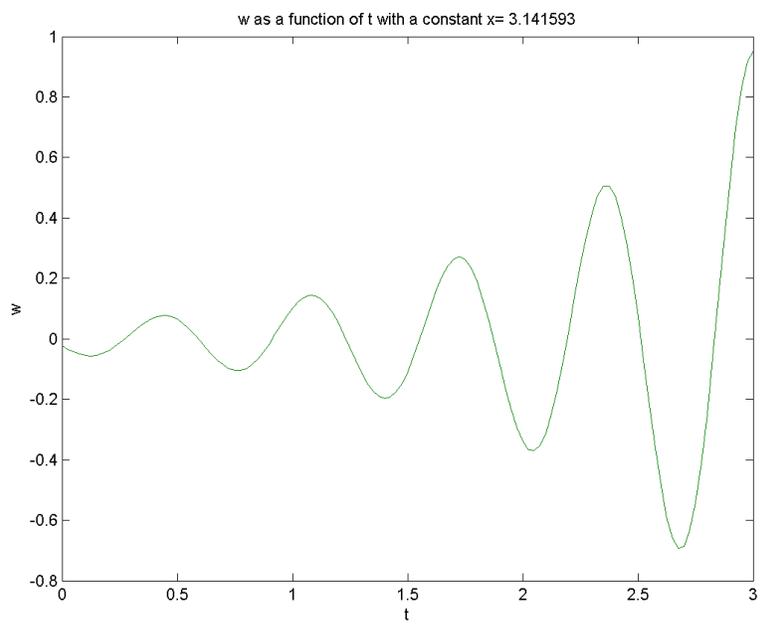

Velocities were also plotted for x with t held constant for t = 0, t =2 t, = 4

t = 0

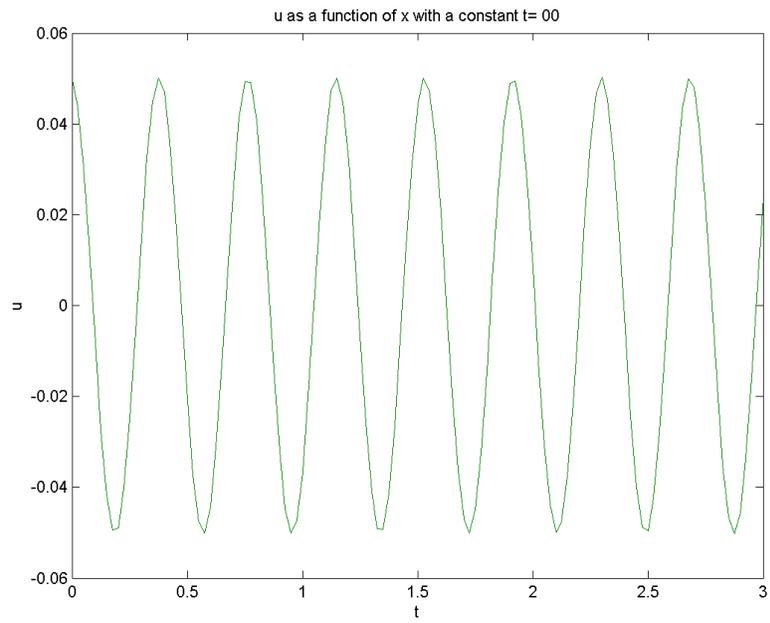

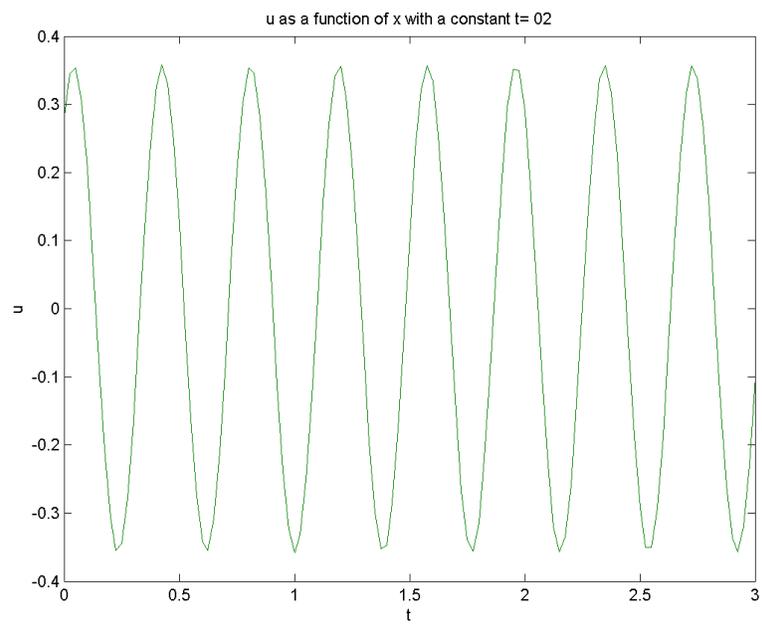

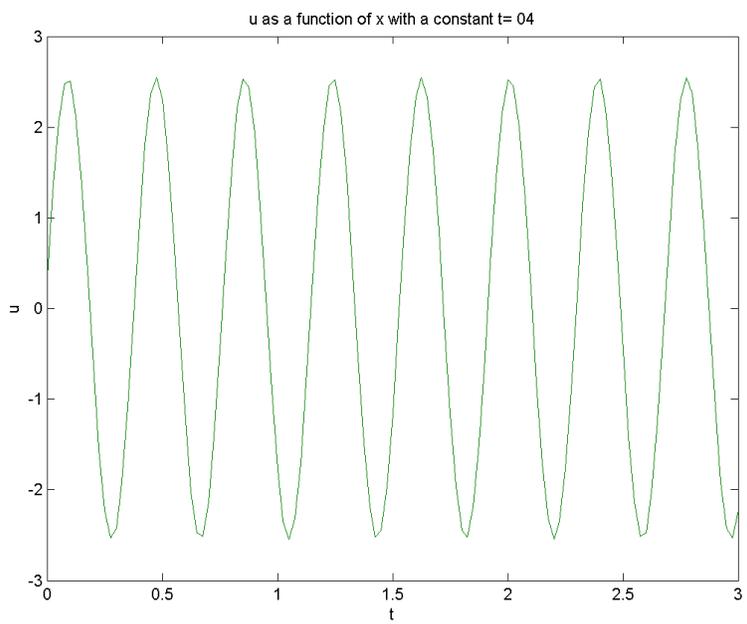

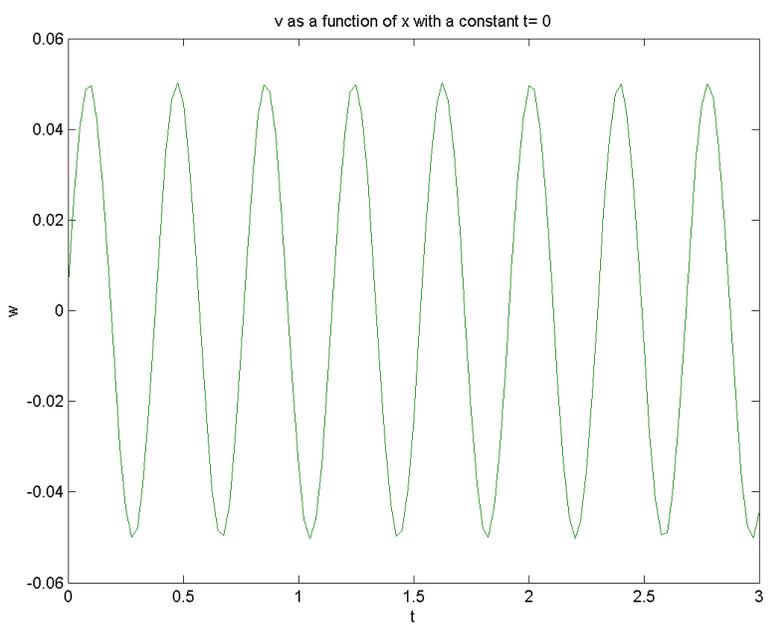

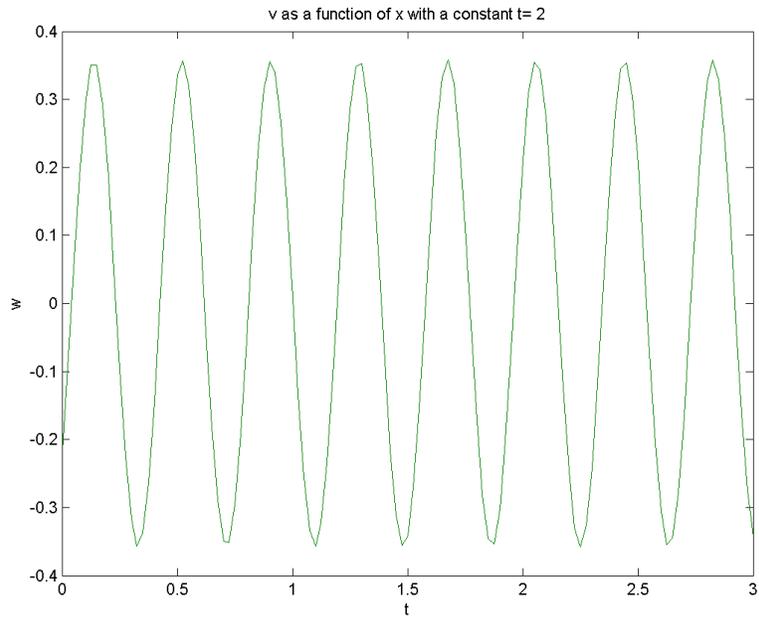

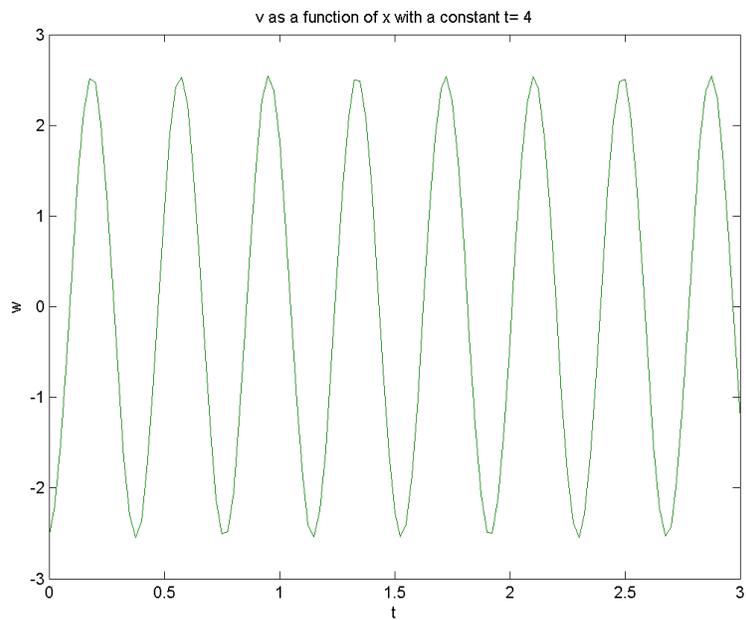

Velocity in both the x and y direction can be can be shown as being sinusoidal with similar magnitudes at each maxima and minima.

Lastly $\beta = \dfrac{2\beta_{WnM}}{\omega}$ was plotted in the same way $\beta_{WnM}$ was plotted for each variable using a Surface plot. The plots are similar to Weber and Melsom in magnitude but are sinusoidal in nature.

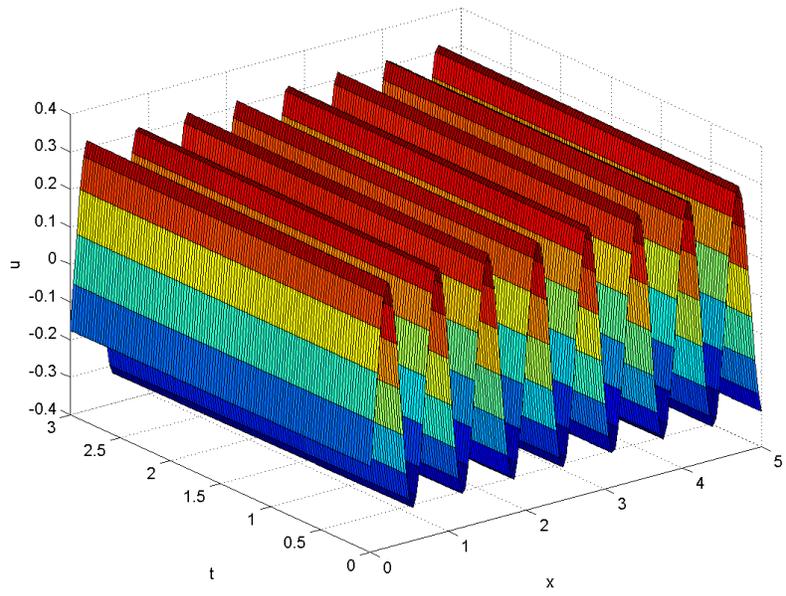

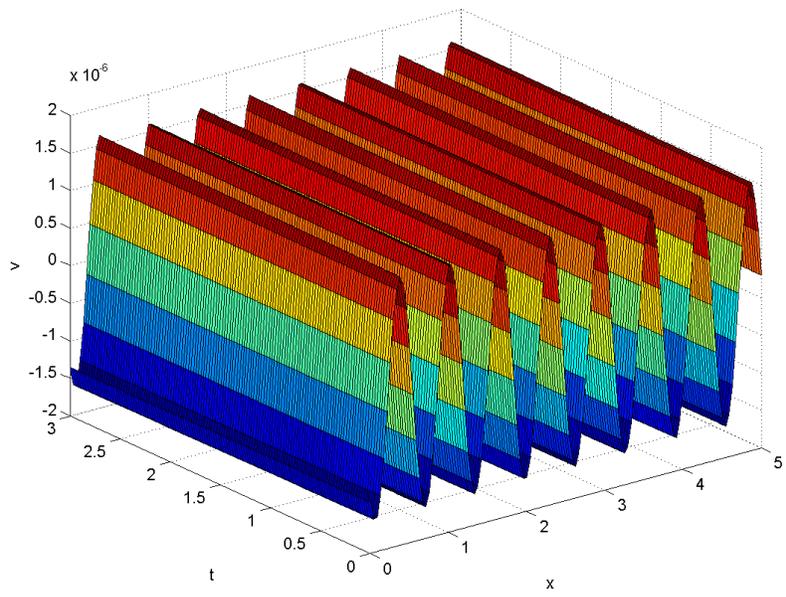

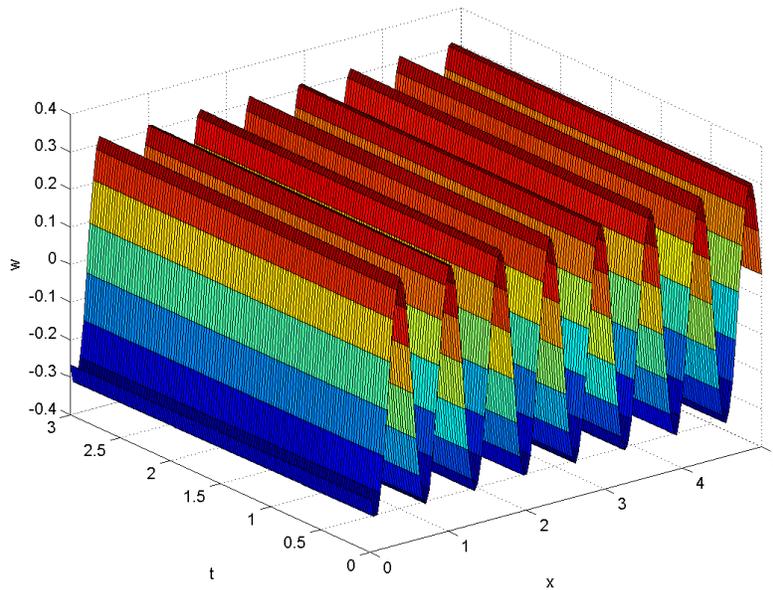

The velocities in this model are sinusoidal with maxima and minima at the same height.

### IV. Conclusion

As with Sajjadi Hunt and Drullion the energy transfer parameter can be shown with as function of wave age, and that due to energy transfer and wave growth modeling velocities show that they are indeed sinusoidal in nature. More research must be done in the transfer of energy into waves causing them to break and release water spray into the air. Hopefully with that it will be able to show the effects of the WISHE model for tropical cyclone development

### V. Acknowledgements

A special thank you and Acknowledgement to Shahrdad Sajjadi for all the help I received in this study. Without him I doubt I could have gotten very far.